\documentclass{appolb}
\usepackage{graphicx,color,amsmath,amssymb,geometry}

\begin{document}
\title{Selected Topics in Quark-Hadron Physics: From Scalar Nonets to Topological Glueballs%
}
\author{Chihiro Sasaki
\address{Institute of Theoretical Physics, University of Wroc\l aw,
plac Maksa Borna 9, PL-50204 Wroc\l aw, Poland}
\address{International Institute for Sustainability with Knotted Chiral Meta Matter (WPI-SKCM$^2$),
Hiroshima University, 1-3-1 Kagamiyama, 739-8531, Higashi-Hiroshima, Hiroshima, Japan}
}
\maketitle
\begin{abstract}
This contribution reviews recent progress in the low-lying scalar mesons and glueballs.
We propose a new classification for the scalar nonet that includes $f_0(980)$ and $a_0(980)$ as the lowest states,
while we identify $f_0(1500)$ as a primary glueball candidate. 
We demonstrate that the production yields of these states in heavy-ion collisions are mutually consistent across statistical, coalescence, and $S$-matrix frameworks.
To investigate their internal structure, we move beyond standard phenomenology by describing glueballs as topological solitons.
This approach yields an energy spectrum in excellent agreement with lattice QCD and experimental data, while interpreting $f_0(2470)$ as a tightly bound glueballonium to explain its anomalously long lifetime.
This non-perturbative framework provides a predictive basis for the future experimental verification of exotic scalar states.
\end{abstract}

\section{Introduction}
The classification of scalar mesons has been one of the long-standing contentious issues in hadron physics~\cite{ParticleDataGroup:2024cfk}. 
Despite the success of the constituent quark model, the scalar sector, comprising the $f_0(980)$, $a_0(980)$, $K^*_0(1430)$, and $f_0(1770)$, exhibits properties that are difficult to reconcile with simple $q\bar{q}$ states. 
Furthermore, the search for the scalar glueball remains a central challenge. 
Whereas various candidates such as $f_0(1500)$ and $f_0(1710)$ have been proposed, a definitive consensus on their nature is still lacking.

In this contribution, we propose a novel classification scheme for the scalar nonet and identify $f_0(1500)$ as a primary glueball candidate~\cite{Yasui:2026vve}.
To examine the validity of this assignment, we investigate their production yields in relativistic heavy-ion collisions (HICs). 
By employing both the statistical model and the coalescence model, we provide robust estimates that show mutual consistency between the two frameworks. 
These results are further supported by an analysis using the S-matrix approach.

To characterize the exotic nature of these states, we go beyond the standard phenomenological approaches: the study of glueballs in terms of topological solitons could shed light on the physics of exotic hadrons from a different perspective. 
Conventional models often treat these particles as point-like objects, leaving their internal structures, such as spatial sizes, shapes, and energy density distributions, largely unexplored.

To address this gap, we employ a semiclassical description within the Skyrme-Faddeev model~\cite{Amari:2025cre}. 
This approach provides a direct parallel to the Skyrme model, where nucleons emerge as topological solitons composed of pion fields,
similarly, in our framework, glueballs are described as solitons (Hopfions) composed of gluonic fields.
By quantizing these Hopfions as rigid bodies, we construct energy spectra that can be directly confronted with experimental data and lattice QCD results. 
This dual approach, combining HIC phenomenology with a topological analysis of internal structure, offers a comprehensive framework for unraveling the nature of glueball.

\section{New scalar nonet}
\subsection{Motivations}

Since the early stages of hadron spectroscopy, light-flavor scalar mesons have been a focal point of intense research. 
They often manifest as {\it exotic} hadronic structures beyond a conventional quark model, including multi-quarks, hadronic molecules, glueballs, and hybrids.
For decades, low-lying scalar mesons, specifically the $f_0(500)$ (iso-singlet), $K^*_0(700)$ (iso-doublet), $f_0(980)$ (iso-singlet), and $a_0(980)$ (iso-triplet), 
have been considered as primary candidates for exotic states. This is motivated by a specific mass hierarchy observed in experiments:
\begin{equation}
m_{f_0(500)} < m_{K^*_0(700)} < m_{f_0(980)} \simeq m_{a_0(980)}\,.
\end{equation}
This mass ordering remains a significant puzzle since it contradicts the predictions of the naive quark model, which treats mesons as $q\bar{q}$ (quark-antiquark) pairs
in a relative P-wave state~\cite{Jaffe:1976ig,Jaffe:1976ih}.
In a conventional $q\bar{q}$ framework, the mass ordering should be dictated by the mass of their constituent quarks based on $SU(3)$ flavor symmetry.
Since the observed mass ordering does not align with these $q\bar{q}$ expectations, several alternative structures have been proposed to resolve the discrepancy:
compact tetra-quarks ($q^2\bar{q}^2$)~\cite{Jaffe:1976ig,Jaffe:1976ih}, extended hadronic molecules ($\pi\pi$, $\pi K$, or $K\bar{K}$)~\cite{Close:2002zu,Amsler:2004ps,Bugg:2004xu,Klempt:2007cp,Pelaez:2015qba,Pelaez:2021dak},
other structures including glueballs and hybrid states.
These models attempt to provide a more accurate theoretical foundation for the unique properties observed in the scalar sector.
Whereas the diquark model requires a significant $\bar{s}s$ component within $f_0(980)$, recent findings from the ALICE experiment in p-Pb collisions suggest a small $\bar{s}s$ fraction,
implying that it may be a conventional meson without hidden strangeness, in contrast to compact tetra-quark or hadronic molecular interpretations~\cite{ALICE:2023cxn}.
This observation highlights the importance of revisiting the flavor structure of scalar mesons, including $f_0(980)$.

The scalar sector below $2$ GeV comprises a variety of states, see Fig.~\ref{fig1}.
\begin{figure}
\begin{center}
\includegraphics[width=10cm]{250206_spectrum_scalar.pdf}
\caption{
Mass spectra of scalar mesons with non-strangeness and strangeness below $2$ GeV.
The new nonet scalar mesons, $f_0(980)$, $a_0(980)$, $K^*_0 (1430)$, and $f_0(1770)$, proposed in this study are indicated by arrows.
The glueball candidate, $f_0(1500)$, is shown by dashed line.
Figure taken from \cite{Yasui:2026vve}.
}.
\label{fig1}
\end{center}
\end{figure}
Although the $f_0(500)$, $K^*_0(700)$, $f_0(980)$, and $a_0(980)$ are traditionally viewed as an $SU(3)$ octet, we re-examine this classification by excluding the $f_0(500)$ and $K^*_0(700)$.
Given their significant decay widths, we argue that these broad resonances should not be categorized alongside the more stable scalar multiplets.
Instead, in Ref.~\cite{Yasui:2026vve}, we propose a new nonet structure composed of $f_0(980)$, $a_0(980)$, $K^*_0(1430)$, and $f_0(1770)$ as in Fig.~\ref{fig1} and Fig.~\ref{fig2} (a).
\begin{figure}
\begin{center}
\includegraphics[width=7cm]{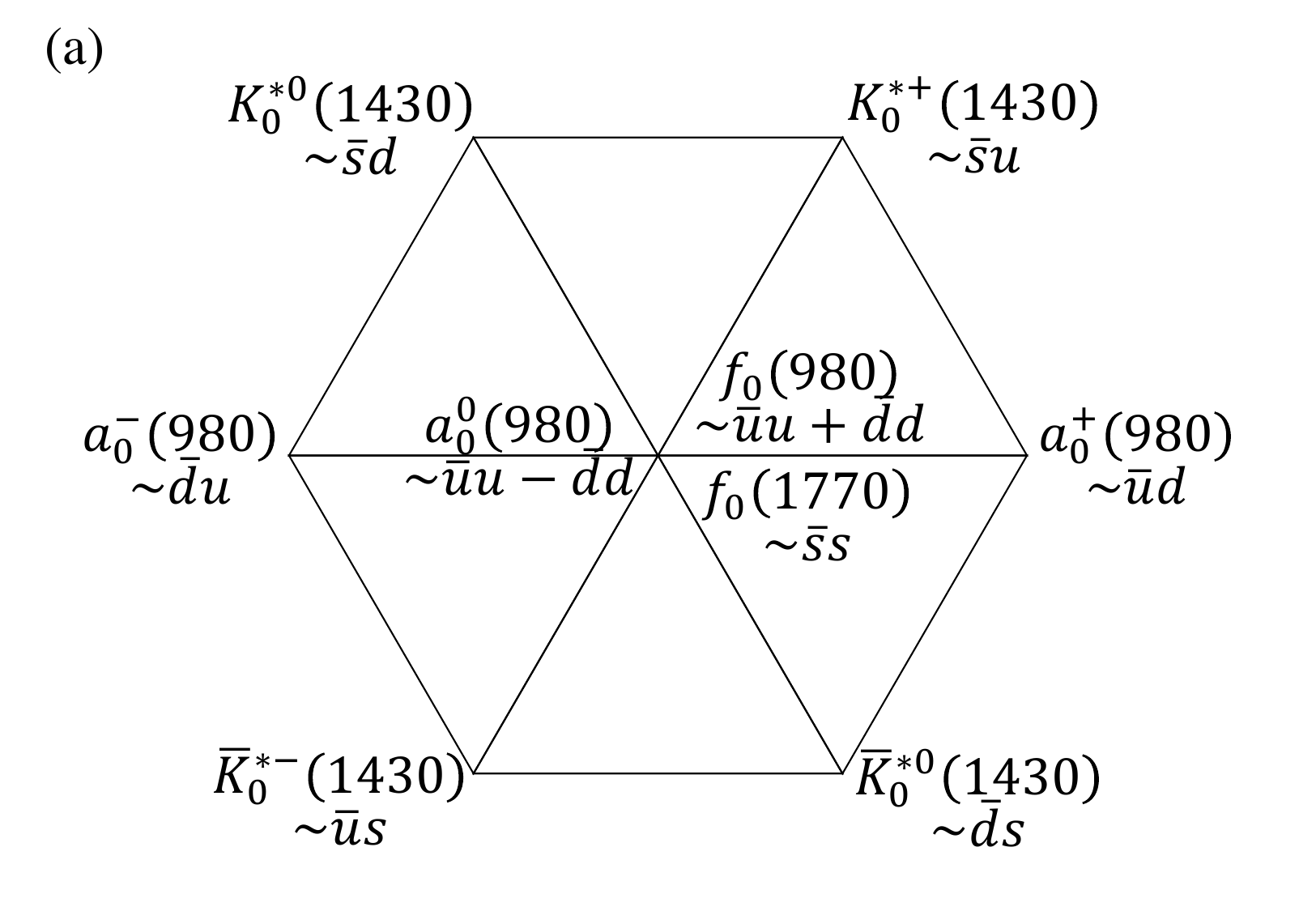}
\includegraphics[width=7cm]{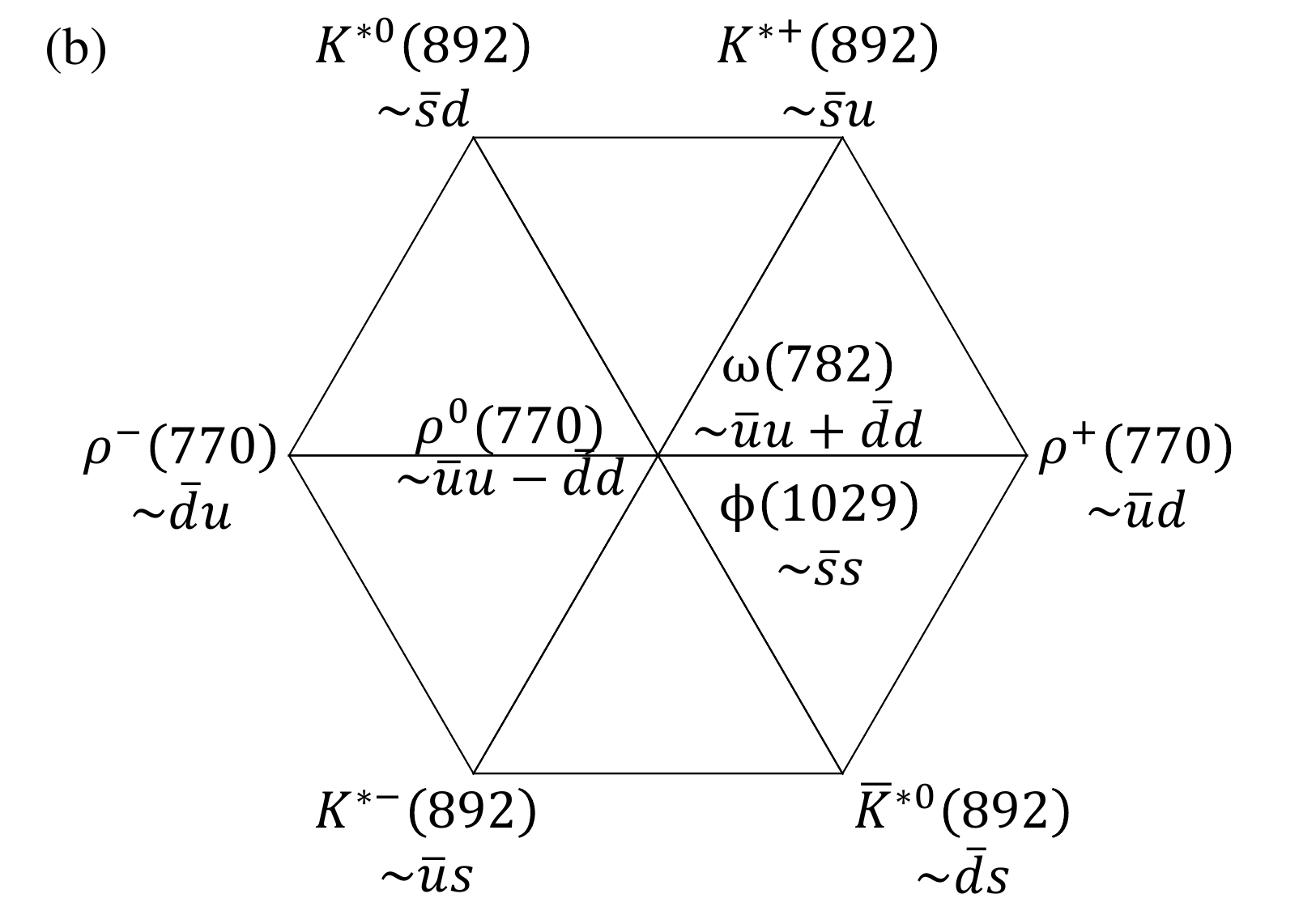}
\caption{
The SU(3) flavor diagrams: (a) new nonet scalar mesons proposed in this study and
(b) ordinary vector mesons.
Figures taken from \cite{Yasui:2026vve}.
}.
\label{fig2}
\end{center}
\end{figure}
The $f_0(980)$ and $a_0(980)$ are regarded as the lowest states in the new nonet.
This classification follows a hierarchy analogous to the well-known vector meson nonet, consisting of $\omega(782)$, $\rho(770)$, $K^*(892)$, and $\phi(1020)$, as shown in Fig.~\ref{fig2} (b).
The $f_0(1710)$ and $a_0(1710)$ are not included in our primary nonet scheme, as they cannot be reconciled with the expected mass ordering alongside $f_0(980)$.
It is far more likely that these resonances, together with $K^*_0(1950)$, form part of a distinct heavier nonet.
Under this assumption, $f_0(2100)$ or $f_0(2200)$ would serve as the $\bar{s}s$ constituent.

\subsection{Production of scalar mesons and glueballs in heavy-ion collisions}

The production mechanisms of the suggested scalar nonet and glueball states can be studied in the context of Heavy-ion collisions (HICs).
To establish a baseline, we first apply the statistical model, subsequently, we incorporate the quark coalescence model to investigate the sensitivity of yields to the underlying structures of mesons.

We begin with employing the statistical model~\cite{Andronic:2005yp}, which determines hadron yields based on the thermal distribution function at a given temperature.
The model is characterized by its simplicity, requiring only the hadron mass and the degree of degeneracy as fundamental inputs.
Nevertheless, it has demonstrated remarkable success in describing the experimental yields of various hadrons and nuclei, including resonances, produced in HICs.
Therefore, utilizing the statistical model serves as a reliable baseline for evaluating our proposed scalar nonet and glueball states.
The parameters used in this work follow Refs.~\cite{Andronic:2012dm,Stachel:2013zma}, with hadronization temperature $T_H = 162$ ($156$) MeV, volume $V_H = 2100$ ($5380$) fm$^3$,
baryon chemical potential $\mu_B = 24 (0)$ MeV and strangeness chemical potential $\mu_S = 10 (0)$ MeV chosen for RHIC (LHC).
As noted in~\cite{ExHIC:2017smd}, these parameters account for the isentropic expansion of the medium, linking the temperature and volume.
This parameter set is well validated, as it successfully explains the yields of standard hadrons like $\rho(770)$, $\phi(1020)$, and $\Omega(1670)$.

The internal structures of final-state hadrons are addressed through the coalescence model, where the $q\bar{q}$ and $gg$ wave functions are essential components.
By convoluting these wave functions with the thermal distribution functions of quarks and gluons, we determine the production yields.
Following the methodology in \cite{ExHIC:2010gcb,ExHIC:2011say,ExHIC:2017smd}, we adopt a harmonic oscillator potential within a non-relativistic Schr\"{o}dinger equation to describe the confined bound states.
The glueball is specifically modeled as a two-gluon system with an effective dynamical mass $m_g$.
Following Ref.~\cite{ExHIC:2017smd}, we determine the critical parameters $T_C$ and $V_C$ through entropy conservation, considering two specific scenarios:
\begin{itemize}
\item
{\bf Scenario 1}:
Assuming a continuous crossover transition where $T_C = T_H$ and $V_C = V_H$. Here, the statistical and coalescence models are expected to yield identical hadron numbers.
\item
{\bf Scenario 2}: $T_C$ is adjusted so that the coalescence yields at the critical point equate to the statistical yields at $T_H$.
\end{itemize}
The spatial extension of a hadronic wave function is governed by the oscillator frequency $\omega_h$.
We calibrate $\omega_h$ such that the coalescence model yields for non-exotic hadrons match the predictions of the statistical model.
This matching condition is grounded in the fact that the statistical model serves as a reliable prototype for hadron production in HICs, given its consistent agreement with experimental data.

We further estimate yields using the S-matrix formulation, where the thermal properties of interacting hadrons are determined by scattering phase shifts.
Here, the energy derivative of the phase shift functions as an effective density of states, allowing the full interaction dynamics to be reflected in the resulting statistical yields.

\subsection{Hadron yields}

The calculated yields for the proposed scalar nonet mesons fall predominantly within the range of $10^{-1}$ to $10^{0}$.
These values are comparable to the yields of the reference particles, $\rho(770)$, $\phi(1020)$, and $\Omega(1672)$, despite their distinct internal angular momentum structures.
Consequently, our results suggest that the experimental detection of the nonet scalar mesons, specifically $f_0(980)$, $a_0(980)$, $K^*_0(1430)$, and $f_0(1770)$, is highly feasible in future measurements at RHIC and LHC.

The yields for $f_0(1500)$ are calculated, considering various internal structures: $gg$ ($S$-wave), $n\bar{n}$ ($P$-wave), and $n^2\bar{n}^2$ ($S$-wave).
We find that the production rates are strongly dependent on the assumed substructure.
The $n\bar{n}$ configuration exhibits a higher yield than the $gg$ state, as the lightness of the $u, d$ quarks compensates for the $P$-wave centrifugal barrier.
On the other hand, the tetra-quark ($n^2\bar{n}^2$) yield is suppressed relative to the glueball ($gg$) state.
This reflects the reduced formation probability of a four-body bound state compared to a two-body system, assuming identical angular momentum.
Detailed numerical results can be found in Ref.~\cite{Yasui:2026vve}.

\begin{figure}
\begin{center}
\includegraphics[width=14cm]{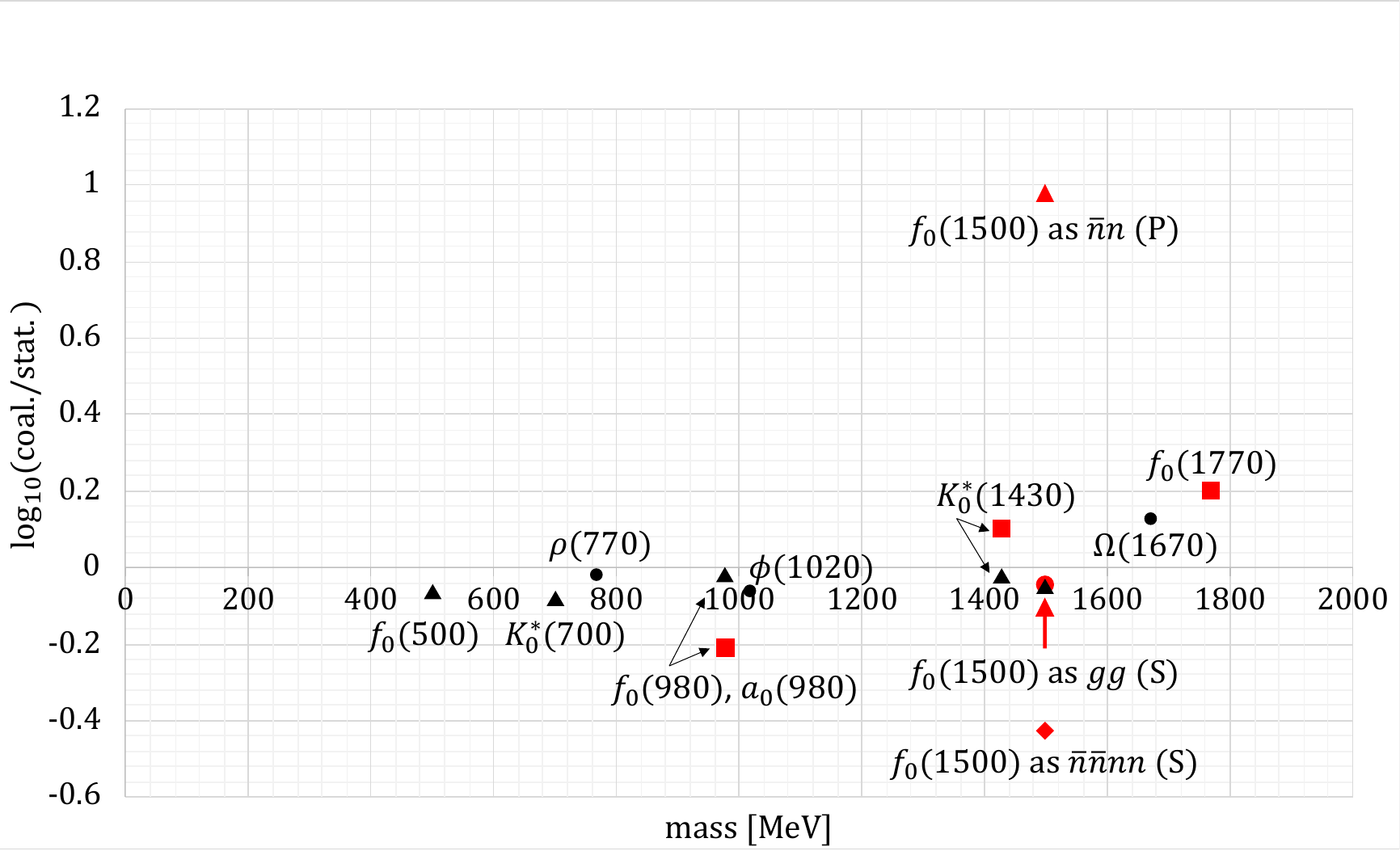}
\caption{
The ratios of the hadron yields from the statistical model and the coalescence model for the RHIC parameter set in Scenario 1.
The normal hadrons are indicated by the black dots, while the new nonet mesons by the red squares.
The ratios from the S-matrix formulation are shown by black-triangles.
Figure taken from \cite{Yasui:2026vve}.
}.
\label{fig3}
\end{center}
\end{figure}

To clarify the correspondence between the coalescence and statistical models, we examine their yield ratio, coal./stat.
Consistent with previous findings, Figure~\ref{fig3} shows that for RHIC parameters under Scenario 1, the ratios for normal hadrons ($\rho, \phi$, and $\Omega$) are approximately one.
This indicates that the two models are in good agreement for conventional states. A similar trend is observed when applying LHC parameters.
The yield ratios for the new nonet scalar mesons are also shown in the same figure:
While these ratios are around unity, they exhibit a slightly more pronounced deviation than those of conventional hadrons.
It is noted that the yield ratio for $f_0(1500)$ under the $S$-wave $gg$ assignment is also close to one, mirroring the behavior of both normal hadrons and the new scalar nonet.
In contrast, the $P$-wave $n\bar{n}$ and $S$-wave $n^2\bar{n}^2$ configurations for $f_0(1500)$ result in anomalous ratios with significant deviations from unity.
Consequently, we conclude that the $S$-wave $gg$ structure is a more plausible interpretation for $f_0(1500)$ than the alternative ones.

\section{Glueballs as topological solitons}
\subsection{The Skyrme-Faddeev model}

Recent high-energy experiments by the BESIII collaboration~\cite{BESIII:2023wfi} have reported the observation of a prominent glueball candidate, $X(2370)$, following earlier attempts~\cite{Crede:2008vw}. The measured mass, ranging between $2.3$ and $2.6$ GeV, aligns closely with predictions 
from both lattice QCD (LQCD)~\cite{Morningstar:1999rf} and holographic QCD models~\cite{Brower:2000rp,Colangelo:2007pt}.
Despite the available LQCD date~\cite{Morningstar:1999rf,Morningstar:2024vjk,Michael:1988jr,Bali:1993fb,Sexton:1995kd,Gregory:2012hu,Athenodorou:2020ani}
and other theoretical approaches for investigating glueball properties~\cite{Isgur:1984bm,Jaffe:1975fd,Barnes:1981ac,Cornwall:1982zn,Schechter:1980ak,Gomm:1985ut,Brunner:2015oqa,Brunner:2015yha},
their definitive experimental verification remains exceptionally challenging.
These experimental challenges may be overcome by gaining deeper insights into the internal structures of glueballs, including their shapes and energy density distributions, alongside their masses and production/decay properties.
Such characteristics are directly accessible via the Hopfion approach~\cite{Faddeev:1996zj,Faddeev:1998eq,Faddeev:2003aw,Kondo:2006sa}, which semiclassically describes glueballs as Hopfions or knot solitons.
This framework treats glueballs as topological solitons composed of gluons, in parallel to the Skyrme model~\cite{Skyrme:1961vq,Skyrme:1962vh}, where nucleons emerge as Skyrmions (topological solitons composed of pions).

In Ref.~\cite{Amari:2025cre}, we developed a Hopfion-based description of glueballs by performing a comprehensive comparison between the energy spectra derived from quantized Hopfions and the results obtained from both experimental data and lattice QCD.
In the Skyrme-Faddeev approach~\cite{Faddeev:1996zj,Faddeev:1998eq,Shabanov:1999xy,Shabanov:1999uv}, SU(2) Yang-Mills theory is mapped onto an O(3) nonlinear sigma model featuring a four-derivative stability term.
The dynamics are governed by the Lagrangian density,
\begin{equation}
{\mathcal L}
= \frac{\kappa^2}{4} \text{Tr}(\partial_\mu {n} \partial^\mu {n}) + \frac{1}{32e^2} \text{Tr}[\partial_\mu {n}, \partial_\nu {n}]^2\,,
\end{equation}
where the target space is parameterized by a unit vector $\mathbf{n}$ on the two-sphere $S^2$ via the relation ${n} = \boldsymbol{\tau} \cdot \mathbf{n}$ with a vector whose components are Pauli matrices.
The model is characterized by two fundamental constants: $\kappa$, which has the dimension of mass, and the dimensionless coupling $e$.
The Skyrme-Faddeev model admits stable and finite-energy solitons called Hopfions.
These states are classified by their Hopf charge $Q$, an integer invariant reflecting the topological mapping $\pi_3(S^2) \simeq \mathbb{Z}$.
For axially symmetric configurations, the Hopf charge $Q$ is given by the product of two integers, $Q = \ell m^2$,
where $m$ and $\ell$ denote the winding numbers associated with the structure and internal phase of the soliton, respectively.
In what follows, we focus on low-lying glueball states and, accordingly, restrict our analysis to Hopfions with topological charges $Q = 1$ and $Q = 2$.

Hopfion energy spectra can be constructed via collective coordinate quantization, treating the soliton as a rigid body.
Similar to the quantization of Skyrmions in the context of baryons, this method promotes the parameters associated with symmetries of the static energy.
The dynamical system is then quantized using standard canonical procedures.
Technical details can be found in Ref.~\cite{Amari:2025cre}.

Quantized Hopfion energies follow the relation, 
\begin{equation}
E
= M_{\text{cl}}+\frac{J(J+1)}{2V_{11}}+\left(\frac{1}{U_{33}}-\frac{\ell^{2}}{V_{11}}\right)\frac{K_{3}^{2}}{2}\,,
\end{equation}
where $M_{\rm cl}$ is the static mass, $U_{jk}$ and $V_{jk}$ are the inertial tensors,
and $J$ and $K_3$ serve as the spin and isospin-like quantum numbers, respectively. 
By fixing the topological winding numbers $(\ell, m)$ and the quantum numbers $(J, |K_3|)$, the mass of each state can be evaluated for any specified coupling constants.

\subsection{Energy spectra}

To evaluate the Hopfion energy spectrum, we determine the coupling constants $\kappa/e$ and $\kappa e^3$ by aligning the model with established scalar and tensor mesons ($J^P = 0^+, 2^+$):
By identifying the $Q=1$ Hopfion with $f_0(1500)$, we obtain $\kappa/e = 5.54$ MeV. This single parameter successfully reproduces the mass of $f_0(2470)$ via the $Q=2$ configuration.
Additionally, the model predicts a currently unobserved $Q=2$ state at $2814$ MeV, providing a target for future experimental verification.
Incorporating quantum corrections, we fix $\kappa e^3 = 88.5$ GeV by identifying the lowest $Q=1$ state with the tensor glueball candidate $f_J(2220)$.
This leads to a triplet structure for $f_2(2220)$, $f_2(2300)$, and $f_2(2340)$, and all of which are viable tensor glueball candidates.
Furthermore, the model predicts a unique $Q=2$ doublet, representing a genuine signature of the Hopfion approach to be assessed in future experiments.

\begin{figure}
\begin{center}
\includegraphics[width=14cm]{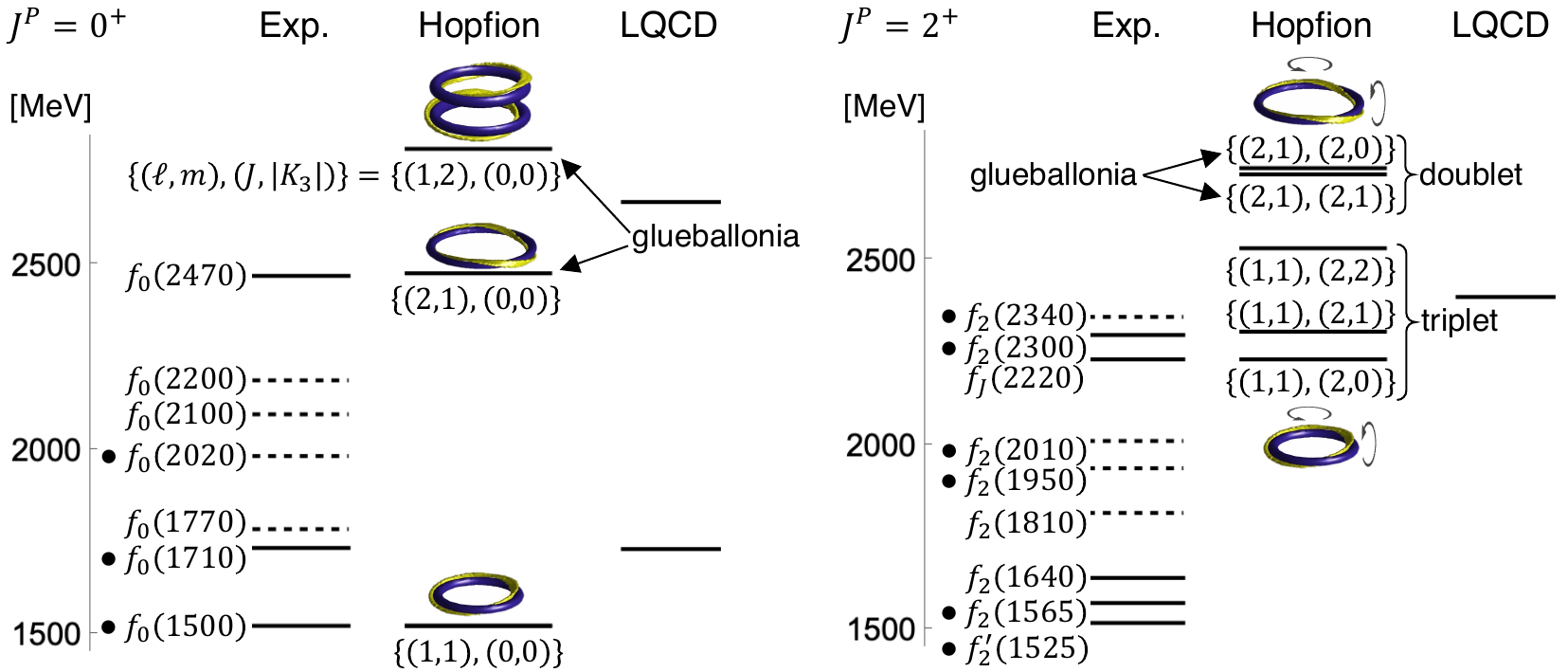}
\caption{
Comparison between the Hopfion energy spectra and the mass spectra of light unflavored mesons listed in PDG and glueballs predicted by LQCD~\cite{Morningstar:1999rf}.
The black dots indicate established particles.
Figure taken from \cite{Amari:2025cre}.
}.
\label{fig4}
\end{center}
\end{figure}

We provide a summary of the Hopfion energies in Fig.~\ref{fig4} illustrating their alignment with PDG scalar/tensor mesons and LQCD glueball candidates~\cite{Morningstar:1999rf}.
We emphasize that the masses of $f_0(2470)$, $f_2(2300)$, and $f_2(2340)$ emerge naturally within this framework without fine-tuning. 
This systematic consistency underscores the robustness of the model, especially as other parameter configurations failed to yield satisfactory results.

In the Hopfion model, heavier scalar glueballs emerge as composite {\it glueballonia} states~\cite{Giacosa:2021brl}. 
This arises from $Q=N$ Hopfions being interpreted as bound systems of $N$ $Q=1$ units. 
We identify the $f_0(2470)$ as a tightly bound $Q=2$ state with binding energy of $577$ MeV, while predicting a novel more loosely bound exotic-state at 2814 MeV with binding energy of $230$ MeV.

We note that the predictive power of this model remains stable even if $f_0(1710)$ is adopted as the primary glueball candidate instead of $f_0(1500)$.
Using $f_0(1710)$ leads to $\kappa/e = 6.31$ MeV and a spectrum that aligns well with lattice QCD benchmarks~\cite{Morningstar:1999rf}.
As discussed in \cite{Amari:2025cre}, these results, assuming also several other possibilities, demonstrate the robustness of our qualitative conclusions
regardless of the initial input.

\section{Summary}
Based on a new scalar nonet and $f_0(1500)$ as a glueball~\cite{Yasui:2026vve}, 
we have provided a comprehensive analysis of their production yields in HICs.
The robustness of our results is demonstrated by the agreement between the statistical and coalescence models, as well as the S-matrix formulation. 
Such theoretical consistency underscores the importance of future experimental searches to validate our classification of the scalar sector.

While the statistical and coalescence models provide insights into production yields, a deeper understanding of the dynamical nature of glueballs is required. 
To this end, shifting from conventional frameworks to the study of glueball production in terms of topological solitons offers a distinct perspective on exotic hadron physics. We have studied low-lying scalar and tensor glueballs in a Hopfion approach~\cite{Amari:2025cre}, 
validated against current experimental data and LQCD predictions.
A key finding is the interpretation of $f_0(2470)$ as a tightly bound glueballonium, contrasting with a newly predicted, loosely bound scalar state. 
Thus, the Hopfion approach explains the unnaturally long lifetime of $f_0(2470)$.
The experimental detection of the novel glueballonium would serve as a critical verification of our model. 
Furthermore, we identified several other $Q=1$ and $Q=2$ Hopfion states, both known and predicted, which are classified into characteristic multiplet structures dictated by the underlying topology.

The formation of glueballonia from constituent glueballs warrants further investigation through a synergistic approach combining relativistic heavy-ion collisions with the real-time dynamics of Hopfions. 
In addition, exploring quantum transitions between distinct glueball states remains a vital task within the Hopfion framework. 
Future research should also prioritize the coupling of Hopfions to dynamical quarks; such a development is essential for modeling glueball decays into quark-antiquark pairs and enabling a more realistic comparison with experimental observations.

\section*{Acknowledgments}
The author is grateful for fruitful collaboration with Y.~Amari, S.~H.~Lee, P.~M.~Lo, M.~Nitta, K.~Shigaki, S.~Yano and S.~Yasui.
This work is supported in part by the Polish National Science Centre (NCN) under OPUS Grant No. 2022/45/B/ST2/01527 and
the World Premier International Research Center Initiative (WPI) through MEXT, Japan.




\begin{thebibliography}{99}
\bibitem{ParticleDataGroup:2024cfk}
S.~Navas \textit{et al.} [Particle Data Group],
Phys. Rev. D \textbf{110}, no.3, 030001 (2024).

\bibitem{Yasui:2026vve}
S.~Yasui, S.~H.~Lee, P.~M.~Lo and C.~Sasaki,
[arXiv:2603.13764 [hep-ph]].

\bibitem{Amari:2025cre}
Y.~Amari, M.~Nitta, C.~Sasaki, K.~Shigaki, S.~Yano and S.~Yasui,
Phys. Lett. B \textbf{869}, 139805 (2025).


\bibitem{Jaffe:1976ig}
R.~L.~Jaffe,
Phys. Rev. D \textbf{15}, 267 (1977).

\bibitem{Jaffe:1976ih}
R.~L.~Jaffe,
Phys. Rev. D \textbf{15}, 281 (1977).

\bibitem{Close:2002zu}
F.~E.~Close and N.~A.~Tornqvist,
J. Phys. G \textbf{28}, R249-R267 (2002).

\bibitem{Amsler:2004ps}
C.~Amsler and N.~A.~Tornqvist,
Phys. Rept. \textbf{389}, 61-117 (2004).

\bibitem{Bugg:2004xu}
D.~V.~Bugg,
Phys. Rept. \textbf{397}, 257-358 (2004).

\bibitem{Klempt:2007cp}
E.~Klempt and A.~Zaitsev,
Phys. Rept. \textbf{454}, 1-202 (2007).

\bibitem{Pelaez:2015qba}
J.~R.~Pelaez,
Phys. Rept. \textbf{658}, 1 (2016).

\bibitem{Pelaez:2021dak}
J.~R.~Pel{\'a}ez, A.~Rodas and J.~R.~de Elvira,
Eur. Phys. J. ST \textbf{230}, no.6, 1539-1574 (2021).

\bibitem{ALICE:2023cxn}
S.~Acharya \textit{et al.} [ALICE],
Phys. Lett. B \textbf{853}, 138665 (2024).

\bibitem{Andronic:2005yp}
A.~Andronic, P.~Braun-Munzinger and J.~Stachel,
Nucl. Phys. A \textbf{772}, 167-199 (2006).

\bibitem{Andronic:2012dm}
A.~Andronic, P.~Braun-Munzinger, K.~Redlich and J.~Stachel,
Nucl. Phys. A \textbf{904-905}, 535c-538c (2013).

\bibitem{Stachel:2013zma}
J.~Stachel, A.~Andronic, P.~Braun-Munzinger and K.~Redlich,
J. Phys. Conf. Ser. \textbf{509}, 012019 (2014).

\bibitem{ExHIC:2017smd}
S.~Cho \textit{et al.} [ExHIC],
Prog. Part. Nucl. Phys. \textbf{95}, 279-322 (2017).

\bibitem{ExHIC:2010gcb}
S.~Cho \textit{et al.} [ExHIC],
Phys. Rev. Lett. \textbf{106}, 212001 (2011).

\bibitem{ExHIC:2011say}
S.~Cho \textit{et al.} [ExHIC],
Phys. Rev. C \textbf{84}, 064910 (2011).

\bibitem{BESIII:2023wfi}
M.~Ablikim \textit{et al.} [BESIII],
Phys. Rev. Lett. \textbf{132}, no.18, 181901 (2024).

\bibitem{Crede:2008vw}
V.~Crede and C.~A.~Meyer,
Prog. Part. Nucl. Phys. \textbf{63}, 74-116 (2009).

\bibitem{Morningstar:1999rf}
C.~J.~Morningstar and M.~J.~Peardon,
Phys. Rev. D \textbf{60}, 034509 (1999).

\bibitem{Brower:2000rp}
R.~C.~Brower, S.~D.~Mathur and C.~I.~Tan,
Nucl. Phys. B \textbf{587}, 249-276 (2000).

\bibitem{Colangelo:2007pt}
P.~Colangelo, F.~De Fazio, F.~Jugeau and S.~Nicotri,
Phys. Lett. B \textbf{652}, 73-78 (2007).

\bibitem{Morningstar:2024vjk}
C.~Morningstar,
PoS \textbf{LATTICE2024}, 004 (2024).

\bibitem{Michael:1988jr}
C.~Michael and M.~Teper,
Nucl. Phys. B \textbf{314}, 347-362 (1989).

\bibitem{Bali:1993fb}
G.~S.~Bali \textit{et al.} [UKQCD],
Phys. Lett. B \textbf{309}, 378-384 (1993).

\bibitem{Sexton:1995kd}
J.~Sexton, A.~Vaccarino and D.~Weingarten,
Phys. Rev. Lett. \textbf{75}, 4563-4566 (1995).

\bibitem{Gregory:2012hu}
E.~Gregory, A.~Irving, B.~Lucini, C.~McNeile, A.~Rago, C.~Richards and E.~Rinaldi,
JHEP \textbf{10}, 170 (2012).

\bibitem{Athenodorou:2020ani}
A.~Athenodorou and M.~Teper,
JHEP \textbf{11}, 172 (2020).

\bibitem{Isgur:1984bm}
N.~Isgur and J.~E.~Paton,
Phys. Rev. D \textbf{31}, 2910 (1985).

\bibitem{Jaffe:1975fd}
R.~L.~Jaffe and K.~Johnson,
Phys. Lett. B \textbf{60}, 201-204 (1976).

\bibitem{Barnes:1981ac}
T.~Barnes,
Z. Phys. C \textbf{10}, 275 (1981).

\bibitem{Cornwall:1982zn}
J.~M.~Cornwall and A.~Soni,
Phys. Lett. B \textbf{120}, 431 (1983).

\bibitem{Schechter:1980ak}
J.~Schechter,
Phys. Rev. D \textbf{21}, 3393-3400 (1980).

\bibitem{Gomm:1985ut}
R.~Gomm, P.~Jain, R.~Johnson and J.~Schechter,
Phys. Rev. D \textbf{33}, 801 (1986).

\bibitem{Brunner:2015oqa}
F.~Br{\"u}nner, D.~Parganlija and A.~Rebhan,
Phys. Rev. D \textbf{91}, no.10, 106002 (2015).

\bibitem{Brunner:2015yha}
F.~Br{\"u}nner and A.~Rebhan,
Phys. Rev. Lett. \textbf{115}, no.13, 131601 (2015).


\bibitem{Faddeev:1996zj}
L.~D.~Faddeev and A.~J.~Niemi,
Nature \textbf{387}, 58 (1997)

\bibitem{Faddeev:1998eq}
L.~D.~Faddeev and A.~J.~Niemi,
Phys. Rev. Lett. \textbf{82}, 1624-1627 (1999).

\bibitem{Faddeev:2003aw}
L.~Faddeev, A.~J.~Niemi and U.~Wiedner,
Phys. Rev. D \textbf{70}, 114033 (2004).

\bibitem{Kondo:2006sa}
K.~I.~Kondo, A.~Ono, A.~Shibata, T.~Shinohara and T.~Murakami,
J. Phys. A \textbf{39}, 13767-13782 (2006).

\bibitem{Skyrme:1961vq}
T.~H.~R.~Skyrme,
Proc. Roy. Soc. Lond. A \textbf{260}, 127-138 (1961).

\bibitem{Skyrme:1962vh}
T.~H.~R.~Skyrme,
Nucl. Phys. \textbf{31}, 556-569 (1962).

\bibitem{Shabanov:1999xy}
S.~V.~Shabanov,
Phys. Lett. B \textbf{458}, 322-330 (1999).

\bibitem{Shabanov:1999uv}
S.~V.~Shabanov,
Phys. Lett. B \textbf{463}, 263-272 (1999).

\bibitem{Giacosa:2021brl}
F.~Giacosa, A.~Pilloni and E.~Trotti,
Eur. Phys. J. C \textbf{82}, no.5, 487 (2022).

\end{thebibliography}
\end{document}